\documentclass[useAMS,usenatbib]{mnras}
\usepackage{natbib}

\usepackage{graphicx}
\usepackage{epsfig}

\usepackage{txfonts}

\usepackage{hyperref}

\usepackage{float}
\usepackage{color}
\usepackage{lscape,graphicx}
\usepackage{journal_shortcuts}
\usepackage{amssymb}
\usepackage{multirow}
\usepackage{afterpage}
\usepackage{ulem}
\usepackage{threeparttable}
\usepackage{comment}
\usepackage{morefloats}
\usepackage{longtable}
\usepackage{pdflscape}
\usepackage{mdwlist}
\usepackage{upgreek}
\newcommand{\kms}{km~s$^{-1}$}

\newcommand{\unitlum}{erg~s$^{-1}~$}
\newcommand{\mbh}{$M_{\rm BH}$}

\title[Measurement of the BH mass of AGN2]{Detection of faint broad emission lines in type 2 AGN: II. 
On the measurement of the BH mass of type 2 AGN and the unified model}

\begin{document}
\author[F. Onori et al. ]{F.~Onori,$^{1}$$^,$$^{2}$$^,$$^{3}$\thanks{E-mail:f.onori@sron.nl} 
F.~Ricci,$^{3}$
F.~La~Franca,$^{3}$ 
S.~Bianchi,$^{3}$ 
A.~Bongiorno,$^{4}$
M.~Brusa,$^{5}$$^,$$^{6}$
F.~Fiore,$^{4}$
\newauthor
R.~Maiolino,$^{7}$
A.~Marconi,$^{8,9}$ 
E.~Sani,$^{10}$
C.~Vignali$^{5}$$^,$$^{6}$
\\
$^{1}$SRON Netherlands Institute for Space Research, Sorbonnelaan 2, 3584 CA Utrecht, Netherlands\\
$^{2}$Department of Astrophysics/IMAPP, Radboud University, P.O. Box 9010, 6500 GL Nijmegen, the Netherlands\\
$^{3}$Dipartimento di Matematica e Fisica, Universit\`a Roma Tre,
   via della Vasca Navale 84, 00146 Roma, Italy\\
$^{4}$INAF - Osservatorio Astronomico di Roma, via Frascati 33, 00044 Monte Porzio Catone, Italy\\
$^{5}$Dipartimento di Fisica e Astronomia, Universit\`a di Bologna, viale Berti Pichat 6/2, 40127 Bologna, Italy\\
$^{6}$INAF - Osservatorio Astronomico di Bologna, via Ranzani 1, 40127 Bologna, Italy\\
$^{7}$Cavendish Laboratory, University of Cambridge, 19 J. J. Thomson Ave., Cambridge CB3 0HE, UK\\
$^{8}$Dipartimento di Fisica e Astronomia, Universit\`a degli Studi di Firenze, Via G. Sansone 1, 50019 Sesto Fiorentino, Italy\\
$^{9}$INAF - Osservatorio Astronomico di Arcetri, Largo E. Fermi 5, 50125 Firenze, Italy\\
$^{10}$European Southern Observatory, Alonso de Cordova 3107, Casilla 19, Santiago 19001, Chile}

\date{Accepted 2017 Month day. Received 2017 Month day; in original form 2016 October 29}

\pagerange{\pageref{firstpage}--\pageref{lastpage}} \pubyear{2016}

\maketitle

\label{firstpage}

\begin{abstract}

We report the virial measurements of the BH mass of a sample of 17 type 2 AGN, drawn from the 
{\it Swift}/BAT 70-month 14-195 keV hard X-ray catalogue, where a faint BLR component has been measured via deep NIR (0.8-2.5 $\mu$m) spectroscopy. 
We compared the type 2 AGN with a control sample of 33 type 1 AGN. We find that the type 2 AGN BH masses span the 5$<$ log($M_{\rm BH}$ /$M_{\sun}$) $< $7.5 range, with an
average log(\mbh /M$_{\sun}$) = 6.7, which is $\sim$ 0.8 dex  smaller than found for type 1 AGN. 
If type 1 and type 2 AGN of the same X-ray luminosity log($L_{\rm 14-195}$/erg s$^{-1}$) $\sim$ 43.5  are compared, type 2 AGN have 0.5 dex smaller BH masses than type 1 AGN. Although based on few tens of objects, this result disagrees with the standard AGN unification scenarios in which type 1 and type 2 AGN are the same objects observed along different viewing angles with respect to  a toroidal absorbing material.

\end{abstract}

\begin{keywords}
galaxies: active --- infrared: galaxies --- quasars: emission lines --- quasars: supermassive black hole
\end{keywords}

\section{Introduction}

In the last decade, by using hard X-ray (2-10 keV) selected Active Galactic Nuclei (AGN) samples, it has been possible to accurately derive the AGN luminosity function up to $z\sim 6$ \citep[e.g.][]{ueda14}. Moreover, by using virial based techniques in the optical band on samples of broad line, type 1,  AGN (AGN1), it has been possible to  estimate the super massive Black Hole (BH) mass function \citep[e.g.][]{greene07a,kelly10, schulze15}.
However, the BH mass (\mbh ) measurements are affected by several selection biases against the narrow line, type 2, AGN (AGN2), or low luminosity AGN, where the Broad Line Region (BLR) is not visible in the rest-frame optical band because of either dust absorption or dilution by the host galaxy spectra \citep[see e.g.][]{baldassare16}.

According to the original standard unified model \citep{antonucci93}, the different observational classes of AGN (AGN1 and AGN2) are believed to be the same kind of objects observed under different conditions (e.g. different orientations of the observer with respect to a dusty torus). In the framework of the AGN phenomenon and co-evolution, this implies that AGN with the same luminosity should share, on average, the same properties (e.g same masses, same accretion rates and, then, same Eddington ratios $\lambda_{\rm Edd}= L_{\rm bol}/L_{\rm Edd}$).   

Nevertheless, nowadays there is growing evidence that AGN1 and AGN2 could belong to intrinsically different populations \citep[see e.g.][]{elitzur12,lanzuisi15}, having, on average, different luminosities \citep[lower for AGN2;][]{ueda03,lafranca05,ueda14},
different accretion rates \citep[smaller for AGN2,][]{winter10,lusso12}, 
different host galaxy properties (more late type for AGN2), different clustering, environment and halo mass properties
\citep{allevato14}.
 
The observed difference in the luminosity distributions of AGN1 and AGN2 could however still comply
with an orientation-based unified model in which the torus opening angle (or the absorbing material covering factor) depends on luminosity.
In this scenario many of the observed differences between the AGN1 and the AGN2 population can be attributed to selection effects.
On the contrary, if a difference
is measured in the average \mbh\ (or host halo mass and clustering properties) of AGN1 and AGN2 {\it sharing the same intrinsic (corrected for absorption) luminosity}, then AGN1 and AGN2 should be intrinsically different objects and the unified model should be revised.

We have therefore started a project aimed at measuring the BH mass in AGN2 \citep{lafranca15,lafranca16,Ricci16b,onori16a}.
In those few studies where AGN2 BH masses have been derived \citep[e.g.][from SDSS]{heckman04}, the authors used the \mbh -host 
scaling relations that have not yet been proven to hold for such a population \citep[see][]{graham08, kormendy11}. 
Several studies have shown that many AGN2 exhibit faint broad line components if observed with high ($\ge$20) S/N in the near-infrared (NIR; 0.8-2.5 $\mu$m), where the dust absorption is less severe than in the optical \citep{Veilleux97,riffel06,cai10}. Moreover observation in the NIR of AGN1, whose \mbh\ were measured using reverberation mapping techniques \citep[RM;][]{blandford82,peterson93}, have demonstrated that the virial method to measure $M_{\rm BH}$ can be efficiently used with the NIR Pa$\alpha$ $\lambda$1.875 $\mu$m and Pa$\beta$ $\lambda$1.282 $\mu$m lines \citep{kim10, landt11a}.

Following the above studies and,  as it is observed for the
optical continuum luminosity, the empirical relation between the X-ray luminosity and
the dimension of the BLR \citep[$R_{\rm BLR} \propto \sqrt{L}$;][]{maiolino07,greene10b},  \citet{Ricci16b} have calibrated new virial relations between the $FWHM_{\rm NIR}$ of the most relevant NIR emission lines (Pa$\alpha$, Pa$\beta$, HeI $\lambda$1.083 $\mu$m) and the intrinsic hard X-ray luminosity, $L_{\rm X}$, of the type $M_{\rm BH}\propto FWHM_{\rm NIR}^2 L_{\rm X}^{0.5}$.
These relations can be used to measure \mbh\ of either AGN2 or obscured and low luminosity  AGN1 \citep[see also][]{lafranca15, lafranca16}.

In this paper we present the measurement for \mbh\ of a sample of  AGN2 selected from the {\it Swift}/BAT 70-month catalogue \citep{baumgartner13} where a faint BLR component in the NIR emission line was found \citep{onori16a}. We compare the resulting \mbh\ distribution with that
of a control sample of AGN1 selected from the {\it Swift}/BAT 70-month catalogue and whose \mbh\  have been measured via RM techniques.
We adopt a $\Omega_{\rm m} = 0.3$, $\Omega_\Lambda = 0.7$ and H$_0$ = 70 \kms Mpc$^{-1}$ cosmology. Unless otherwise stated, all the quoted uncertainties are at 68\% confidence level.

\section{Data and samples}
In order to measure the BH mass of the AGN2, we have carried out NIR spectroscopic observations of 41 obscured and intermediate class AGN (type 2, 1.9 and 1.8; all named AGN2 in the following) at redshift $z \lesssim$0.1,  randomly selected from the {\it Swift}/BAT 70-month catalogue. Thanks to the very hard X-ray band (14-195 keV) that has been used to build the catalogue,
the parent AGN sample is free of absorption selection effects up to log$N_H\lesssim 24$ cm$^{-2}$ \citep[see e.g. Figure 11 in][]{burlon11}.
The observations have been carried out in the framework of a systematic study of the AGN2 NIR spectral properties and have been executed using ISAAC/VLT, X-shooter/VLT and LUCI/LBT spectrometers, with a spectral resolution of 65 \kms, 20-78 \kms and 220 \kms, respectively, and reaching an average S/N ratio of $\sim$30 per resolution element. 
A BLR component showing 800 $<$ FWHM $<$ 3500  \kms, significantly larger than the NLR component measured in the same spectra, has been identified in 13 out of 41 AGN2 ($\sim$ 30\% of the total sample).
The data reduction, spectral analysis and line fitting parameters (FWHM and fluxes of the most relevant emission lines) have been published in a companion paper \citep{onori16a}.  
The sample has been extended with 4 AGN2  included in the  {\it Swift}/BAT 70-month catalogue whose FWHM NIR lines, or spectra, were available in the literature. The spectral data, when available, have been fitted using the technique described in \citet{onori16a}.
In Table \ref{table1} we list the FWHM of the BLR components of the HeI and Pa$\beta$ lines of the AGN2 sample (in the following called AGN2/BLR sample).

\begin{figure}
\centering
\includegraphics[width=\columnwidth]{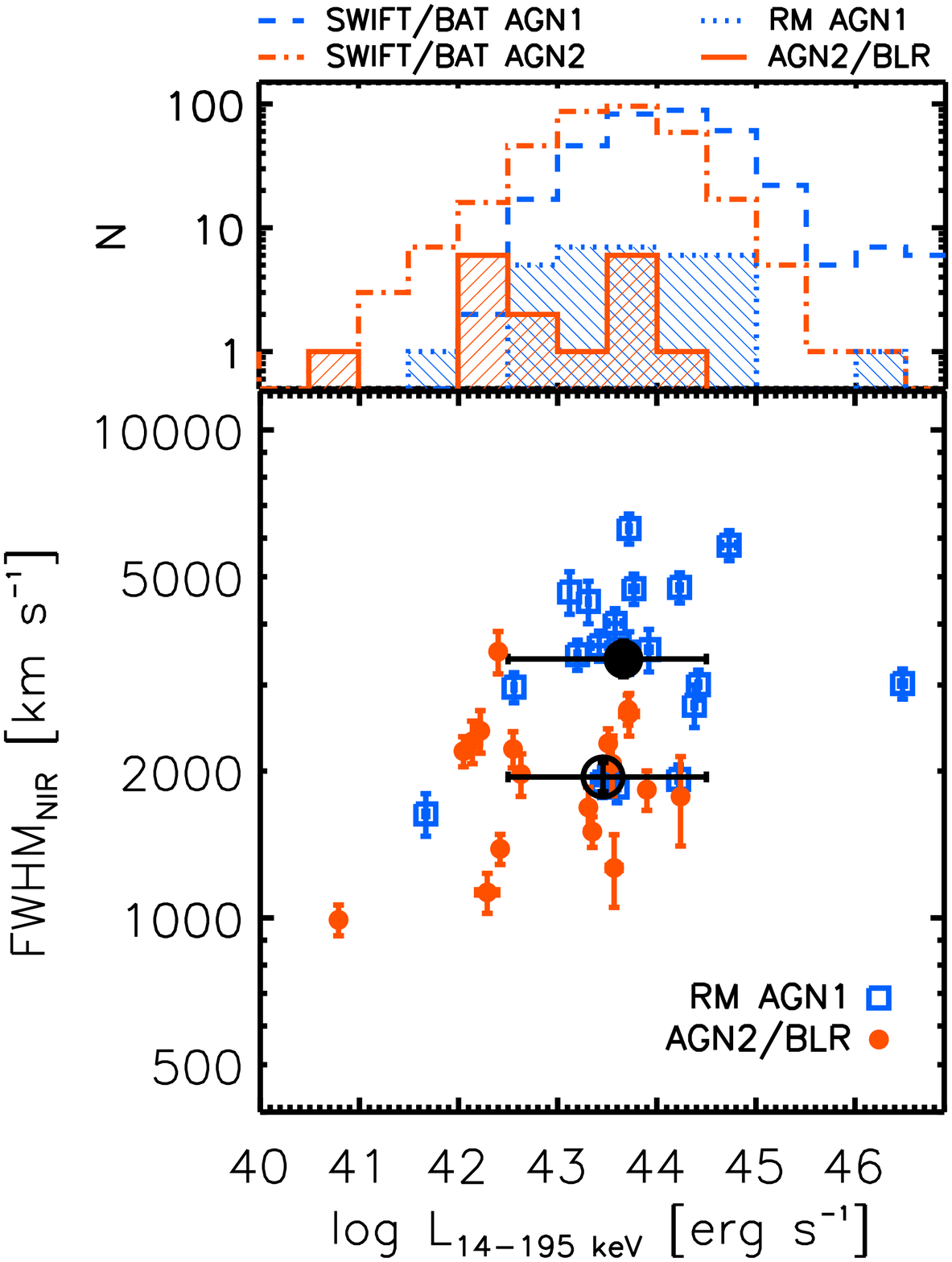}
\caption{{\it Top.} Distribution of $L_{\rm 14-195}$   of the 17 AGN2/BLR, where a BLR component was found (red continuous line), and AGN1 control sample (blue dotted line). The red dot dashed and the blue dashed lines show the distribution of the complete AGN2 and AGN1 samples of the {\it Swift}/BAT 70-month catalogue, respectively.
{\it Bottom.} Average FWHM of the BLR of the NIR lines (Pa$\upbeta$ and HeI) of AGN1 (blue open squares) and  AGN2/BLR (red filled circles) as a function of the intrinsic X-ray luminosity, $L_{\rm 14-195}$  . The
black filled (open) circle shows the FWHM average value of the total AGN1 (AGN2/BLR) sample in the 42.5$<$log($L_{\rm 14-195}$/${\rm erg~ s^{-1}}$)$ <$44.5 luminosity bin and has been plotted at the position of the average log$L_{\rm 14-195}$. }
\label{fig:FWHM}
\end{figure}

In Figure \ref{fig:FWHM} (top) we show the $L_{\rm 14-195}$  distributions of all the AGN2 included in the {\it Swift}/BAT 70-month catalogue (red dot-dashed line) as well as the distributions of the 17 AGN2/BLR (red continuous line). The average X-ray luminosity is log$L_{\rm 14-195}$  =  43.5 with a spread $\sigma$ (log$L_{\rm 14-195}$)=0.9 and  log$L_{\rm 14-195}$  =42.9 \unitlum\ with a spread $\sigma$(log$L_{\rm 14-195}$)=0.9, for the {\it Swift}/BAT AGN2 and the AGN2/BLR, respectively.

In order to build a control sample of AGN1 to be compared with the  results obtained from the AGN2 population we have used 33 AGN1 included in the {\it Swift}/BAT 70-month catalogue and whose \mbh\ have been measured via RM techniques.
This sample includes those 31 AGN1 selected by \citet[][Table 1]{Ricci16b}
plus two additional RM AGN1, namely 3C 390.3 ($\log L_{14-195}$=44.88 \unitlum, 
	$M_{\rm vir}=278^{+24}_{-32}\times 10^6$ M$_\odot$) 
	and Mrk 50 ($\log L_{14-195}$=43.45 \unitlum, 
	$M_{\rm vir}=6.3\pm0.7 \times 10^6$ M$_\odot$) \citep[see Table 1 in][and references therein]{ho14}. 
In Figure \ref{fig:FWHM} (top) we show the $L_{\rm 14-195}$ distribution of all the AGN1 included in the {\it Swift}/BAT 70-month catalogue (blue dashed line) as well as the distribution of the control sample of 33 AGN1 (blue dotted line). The average X-ray luminosity is log$L_{\rm 14-195}$  =  44.3 with a spread $\sigma$ (log$L_{\rm 14-195}$)=1.0 and log$L_{\rm 14-195}$  = 43.8 with a spread $\sigma$ (log$L_{\rm 14-195}$)=1.0, for the {\it Swift}/BAT AGN1 and the control sample of AGN1, respectively.

\begin{table*} 
 \centering\scriptsize
 \begin{minipage}{175mm}
\caption{The sample of  AGN2/BLR and their Black Hole masses}
\label{table1}
\begin{center}
\begin{tabular}{@{}lclccclllccc}
\hline 
                      &       &       &   &   & & \multicolumn{3}{c}{FWHM  }  &      &  &   \\
\cline{7-9} \\
Object Name & $z$    & class &  $A\rm _V$ & W2-W4 & log($L_{\rm 14-195}$) &${\rm He I}$ &${\rm Pa\beta}$ & NIR &log($M\rm _{BH}$) & log$L_{\rm bol}$ & $\lambda_{\rm Edd}$\\ 
                      &       &          & mag & mag & erg s$^{-1}$   &        km s$^{-1}$                &          km s$^{-1}$            & km s$^{-1}$ &        M$_{\odot}$       &   erg s$^{-1}$  &   \\
  \,~ ~ ~~            (1) & (2) & ~ (3)     & (4)& (5)&(6)                     & (7)  & (8)      &  (9)&(10)   & (11) & (12)          \\
\hline
2MASX J05054575--2351139  &  0.0350                   & 2    & ...\phantom{$^a$}   &5.80 & 44.24 & 1772$^{+419}_{-318}$ &  ...                                 &1772$^{+419}_{-318}$  &7.37 $\pm$ 0.18  & 45.50 & 0.849\\
2MASX J18305065+0928414 & 0.0190                    & 2     & ...\phantom{$^a$}   &4.45 & 42.40 &  3513$^{+232}_{-213}$ & ...                                 & 3513$^{+232}_{-213}$  &7.04 $\pm$ 0.09 & 43.09 & 0.009 \\
ESO 234-G-050                      &0.0088                      & 2    & 0.8\phantom{$^a$} &6.30 & 42.29 & 1110$^{+63}_{-59}$       & 1304$^{+381}_{-322}$ &1128$^{+106}_{-106}$ &6.00 $\pm$ 0.10 & 42.98 & 0.076\\  
ESO 374-G-044                      & 0.0284                     & 2    & ...\phantom{$^a$}   &6.96 & 43.57 & 1123$^{+383}_{-221}$  &1412$^{+318}_{-294}$   &1265$^{+215}_{-215}$ &6.74 $\pm$ 0.15 & 44.71 & 0.742\\
MCG -01-24-12                       & 0.0196                     & 2    & ...\phantom{$^a$}   &6.32 & 43.55 & ...                                  &2069$^{+300}_{-280}$    &2069$^{+300}_{-280}$ &7.16 $\pm$ 0.12 & 44.24 & 0.096\\
MCG -05-23-16                       &0.0085                      & 2    & 6.8\phantom{$^a$} &5.84 & 43.51 & 2474$^{+67}_{-64}$     & 2133$^{+93}_{-89}$       &2278$^{+162}_{162}$ &7.22 $\pm$ 0.06 & 44.20 & 0.075\\
 Mrk~~ 1210                           &0.0135                       & 2    & ...\phantom{$^a$}   &6.92& 43.35 & 1305$^{+73}_{-32}$     &1936$^{+118}_{-225}$     &1502$^{+108}_{-108}$& 6.78 $\pm$ 0.06 & 44.36 & 0.303\\ 
 NGC 1052                             &0.0050                       & 2    & 1.5$^a$ &5.12& 42.22 & 2417$^{+143}_{-128}$ &  ...                                                    &2417$^{+143}_{-128}$ &6.63 $\pm$ 0.09 & 42.91 & 0.015\\
 NGC 1365                             &0.0055                       & 1.8 & 5.2\phantom{$^a$} &7.31& 42.63 &  ...                         &1971$^{+85}_{-75}$  &1971$^{+85}_{-75}$ &6.65 $\pm$ 0.09 & 43.32 & 0.037\\
 NGC 2992                             &0.0077                       & 2    & 5.1$^a$ &6.25& 42.55& 3157$^{+586}_{-400}$      &2055$^{+29}_{-30}$   &2218$^{+190}_{-190}$ &6.72 $\pm$ 0.08 & 43.24 & 0.026\\
NGC 4395                              & 0.0013                      & 1.9 & 4.0\phantom{$^a$} &6.32& 40.79& 1332$^{+93}_{-70}$       &\phantom{0}851$^{+29}_{-34}$         &\phantom{0}990$^{+72}_{-72}$  &5.14 $\pm$ 0.07 & 41.48 & 0.017\\
NGC 6221                              &0.0050                       & 2    & 3.2\phantom{$^a$} &7.41& 42.05& 2141$^{+110}_{-141}$        &2256$^{+99}_{-82}$     &2195$^{+155}_{-155}$  &6.46 $\pm$ 0.06 & 42.74 & 0.015\\
NGC 7314                              & 0.0048                      & 1.9 & 4.4\phantom{$^a$} &5.70& 42.42 & 1427$^{+46}_{-38}$       & 1347$^{+46}_{-39}$       &1384$^{+99}_{-99}$  &6.24 $\pm$ 0.06 & 43.11 & 0.058\\
\\
\multicolumn{11}{c}{NIR data taken from the literature} \\
\hline
IRAS F 05189-2524                & 0.0426                      & 2              & ...\phantom{$^a$} &6.26& 43.72 &  ...  & 2619$^b$                    &2619&7.45 $\pm$ 0.10 & 44.41 & 0.073\\
Mrk~~ 348$^c$                       & 0.0150                      & 2/FSRQ   & ...\phantom{$^a$} &5.51& 43.90 & 1917$^{+146}_{-131}$  & 1514$^{+416}_{-319}$         &1830$^{+170}_{-170}$  &7.23 $\pm$ 0.08 & 44.98 & 0.448\\
NGC 1275$^c$                       & 0.0176                      & 2              & ...\phantom{$^a$} &6.91& 43.71 & 2547$^{+20}_{-24}$  &2824$^{+98}_{-85}$          &2671$^{+179}_{-179}$&7.46 $\pm$ 0.06 & 44.40 & 0.069\\
NGC 7465                               & 0.0065                      & 2              & ...\phantom{$^a$} &5.78& 42.14 &   ...     & 2300$^d$                 &2300&6.54 $\pm$ 0.10 & 42.83 & 0.015\\
\hline
\end{tabular}
\end{center}
\small
Notes: (1) AGN name; (2-3) redshift, source classification \citep[from][]{baumgartner13}; (4) optical-NIR extinction; (5) mid-IR 4.6-22 $\mu$m color; (6) 14-195 keV intrinsic luminosity \citep[from][]{baumgartner13}; 
(7)-(8) intrinsic (corrected for instrumental resolution) FWHM of the broad emission line component of {\rm He I} and {\rm Pa$\beta$} from \citet{onori16a}; (9) Average FWHM of the HeI and Pa$\beta$ lines;
(10) \mbh\ derived with the virial relation a3 of Table 4 from \citet{Ricci16b}. A constant virial factor $f= 4.31$ \citep{grier13} has been adopted.
An uncertainty $\epsilon \simeq 0.5$ dex, due to the spread of the population, should also be taken into account \citep{Ricci16b}.
(11)-(12) bolometric luminosity and corresponding Eddington ratio, both calculated adopting the K-correction
of \citet{vasudevan07}.
a) From \citet{burtscher15} b) Measure of the FWHM of the Pa$\alpha$ line from \citet{cai10}. c) The FWHM have been measured applying the fitting procedure described by \citet{onori16a} on the data published by \citet{riffel06}. d) From \citet{ramosalmeida09}.
 \noindent
\end{minipage}
\end{table*}

\section{Comparison between the AGN1 and AGN2 populations}

 As discussed by \citet{landt08,onori16a,Ricci16b},  the most relevant NIR emission lines of the BLR (Pa$\alpha$, Pa$\beta$, HeI $\lambda$1.083 $\mu$m) have, within the errors, the same FWHM. Therefore, a more robust BLR FWHM measure can be obtained using the average width (FWHM$_{NIR}$)  of these lines (when available). In Figure \ref{fig:FWHM} (bottom) we show the FWHM$_{NIR}$ of the BLR, derived using the HeI and Pa$\beta$ lines, as a function
of the intrinsic X-ray luminosity $L_{\rm 14-195}$ for both the AGN1 (blue open squares) and the AGN2/BLR (red filled circles)  samples.
Of the 33 AGN1, only the 20 with NIR emission-line measurements are plotted.
As expected from the studies of the AGN X-ray LF \citep[e.g.][]{ueda03, lafranca05, ueda14} AGN1 have on average larger luminosities than AGN2. 
However, in the  luminosity range where the two distributions overlap, 42.5$<$ log$L_{\rm 14-195}$$<$44.5, AGN1 show significantly larger FWHM than AGN2: $\sim$3400 \kms\  instead of $\sim$1970 \kms (log(FWHM)=3.531$\pm$0.036 and log(FWHM)=3.294$\pm$0.032 for AGN1 and AGN2/BLR, respectively).

\begin{figure}
\centering
\includegraphics[width=\columnwidth]{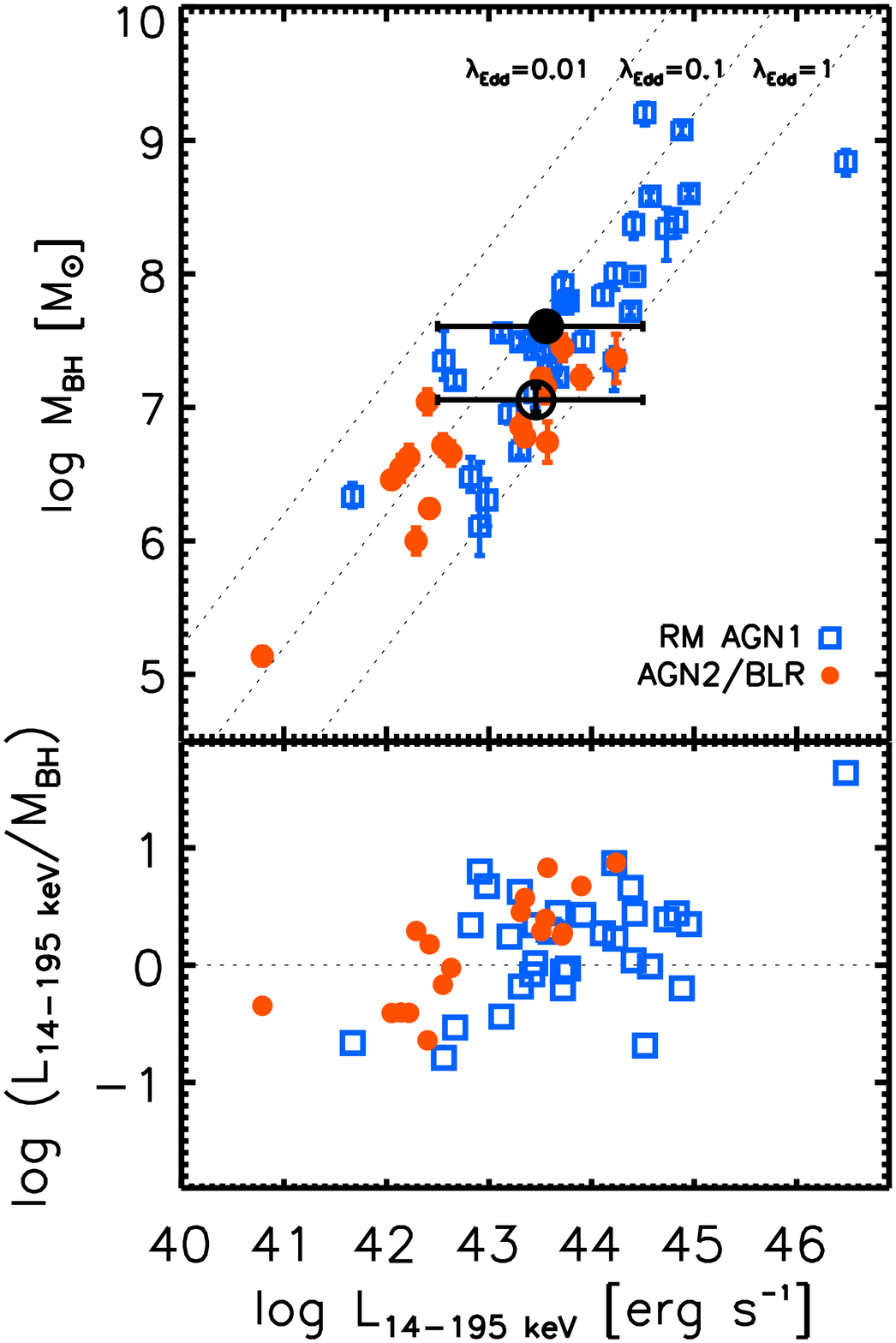}
\caption{{\it Top}. Black hole masses of AGN1 (blue open squares) and AGN2/BLR (red filled circles) as a function of$ L_{\rm 14-195}$ .
	The black filled (open) circle shows the average \mbh\  value of the total AGN1 ( AGN2/BLR) sample in the 42.5$< {\rm log}($L$_{X} /{\rm erg~ s^{-1}}) <$44.5 luminosity bin and has been plotted at the position of the average logL$_{\rm 14-195}$ . The dotted lines show, as a reference, the location of constant $\lambda_{Edd}$, derived assuming a constant $L_{bol}$ = 20$L_{\rm 14-195}$.
	{\it Bottom}.  Ratio between $L_{\rm 14-195}$   and \mbh\ (plus a constant) of the AGN1 (blue open squares) and AGN2/BLR (red filled circles) as a function of $L_{\rm 14-195}$.}
	\label{fig:ML2}
\end{figure}

\begin{figure}
	\centering
	\includegraphics[width=\columnwidth]{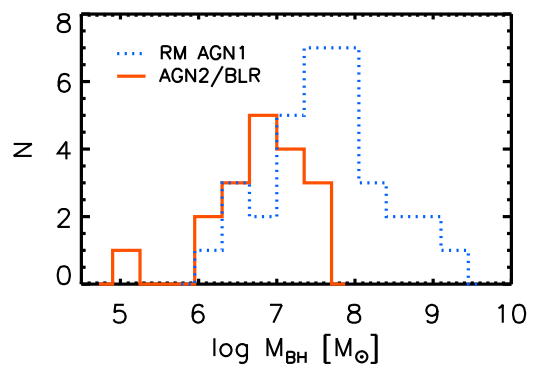}
	\caption{Distribution of $M_{BH}$ of our sample of AGN2/BLR (red continuous line) and of the control sample of AGN1 whose \mbh\ have been measured via RM (blue dotted line). The \mbh\ have been computed assuming $f=4.31$ for both samples \citep[][see text for more details]{grier13}.}
	\label{fig:HistM}
\end{figure}

In order to compute the AGN2 \mbh\ we have used the relation
\begin{equation}
\log\left({M_{BH}\over M_{\sun}}\right) = 7.75 + \log \left[  \left({FWHM_{NIR}\over{10^4~ {\rm km~s^{-1}}}}\right)^2 \left({L_{14-195~{\rm keV}}\over{10^{42}~{\rm erg~s^{-1}}}}\right)^{0.5} \right],
\label{eq2}
\end{equation}

which is based on the measure of the average FWHM observed in the NIR and the hard X-ray 14-195 keV luminosity. The relation has been calibrated by \citet[][see solution a3 in their Table 4]{Ricci16b} assuming a common virial factor $f=4.31$ \citep[][ but see the discussion about this assumption in Sec. \ref{secbias}]{grier13}.
According to the above equation, the measure of \mbh\ depends  on the square root of the luminosity \citep[as typical in the single epoch virial relations,][]{vestergaard02} and the square power of the FWHM. Therefore
the observed narrower (by a  factor $\sim$0.25 dex) FWHM in the AGN2 sample implies (a factor $0.5=2\times 0.25$ dex) smaller \mbh\ for AGN2, if compared with AGN1 of the same luminosity. 

In Figure \ref{fig:ML2} the \mbh\ as a function of $L_{\rm 14-195}$ of the AGN1 (blue open squares) and AGN2/BLR (red filled circles) samples is shown. For the AGN1 sample, \mbh\ has been derived using the RM technique.
As also shown in Figure \ref{fig:HistM}, the \mbh\ of AGN2/BLR (red continuous line)  are typically smaller than AGN1 (blue dotted line). The average (largest) \mbh\ is 
 log(\mbh/M$_{\sun}$) $ \sim 6.8$ ($\sim 7.5$) in the AGN2 sample, while in the control AGN1 sample is
  log(\mbh/M$_{\sun}$) $ \sim 7.6$ ($\sim 9.2$).
In the 42.5$<$ log($L_{\rm 14-195}$ / ${\rm erg~ s^{-1}}$)$<$44.5 luminosity bin the average \mbh\ of the AGN2 sample is $\sim$0.5 dex smaller than that measured in the AGN1 sample (log(\mbh/M$_{\sun}$) $=7.08\pm0.10$ and log(\mbh/M$_{\sun}$) $=7.61\pm0.01$, for AGN2/BLR and AGN1, respectively).
For the sub-sample of 20 AGN1 for which the FWHM$_{NIR}$ measurements are available, the average \mbh\ obtained using  eq. \ref{eq2}, instead of using the RM measurements,  is very similar: log(\mbh/$M_{\sun}$) $=7.63\pm0.01$\footnote{Indeed eq. \ref{eq2} has been calibrated by \citet{Ricci16b} using almost the same RM AGN1 of the control sample (31 out of 37 objects are in common).}.

The above result is also illustrated in Figure \ref{fig:ML2} (bottom), where the  $L_{\rm 14-195}$/\mbh\ ratio (which is a proxy of the Eddington ratio
$\lambda_{\rm Edd}$), plus a constant,  is shown as a function of $L_{\rm 14-195}$.
We computed the Eddington ratio assuming the bolometric correction of \citet{vasudevan07}\footnote{As parameterized by \citet[][eq. 22, in which, because of a typo,  log$L_X$   should read log$\lambda_{\rm Edd}$]{shankar13}. No significant difference in the results was found by using the bolometric correction of \citet{marconi04}.} (see Table \ref{table1}). 
In the overlapping luminosity bin 42.5$<$log($L_{\rm 14-195}$/${\rm erg~ s^{-1}}$)$<$44.5, AGN2 have on average 0.3 dex larger $\lambda_{\rm Edd}$ than AGN1 (log$\lambda_{\rm Edd}\simeq -0.85$ and log$\lambda_{\rm Edd}\simeq -1.15$, for AGN2/BLR and AGN1, respectively).

\section{Analysis of possible selection effects}
\label{secbias}

Although based on few tens of objects, at face value our results imply that AGN2 have on average about 0.25 dex lower BLR FWHM, and
0.5 dex lower \mbh\ (larger $\lambda_{\rm Edd}$) than AGN1 of the same luminosity. In \citet{onori16a} we have investigated whether our FWHM measurements could be affected by some selection biases. No dependence was found in the sample where the BLR was measured  on both the X-ray and NIR fluxes, on the orientation angle of the host galaxy or on the hydrogen column density, $N_{\rm H}$, measured in the X-ray band.

\begin{figure}
\centering
\includegraphics[width=6cm, angle = 90]{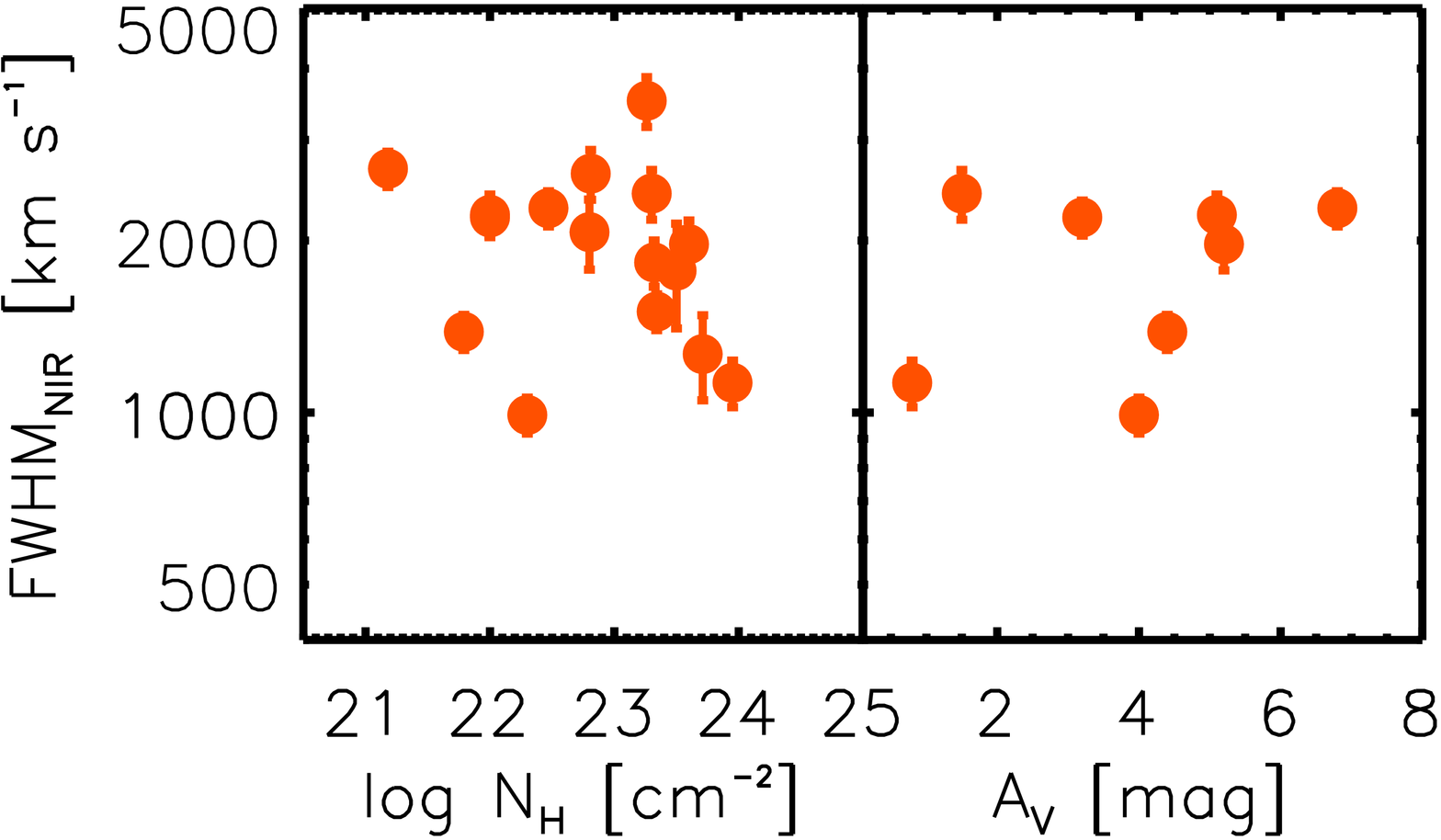}
\caption{{\it Left.} Average FWHM of the BLR components of the NIR HeI and Pa$\beta$ lines of the AGN2/BLR as a function of the Hydrogen column density N$_H$. {\it Right.} Same as the left panel but as a function of the extinction A$_V$.}
\label{fig:FW_NH}
\end{figure}

A possible selection could be originated by the effects of the absorption/reddening
medium along the line of sight, which is obviously present as it is
at the origin of the AGN2 classification, and then large $N_{\rm H}$ values ($>$ 10$^{21}$ cm$^{-2}$ s$^{-1}$) are typically measured in AGN2. One scenario could be that the most central parts of the BLR are embedded in a region of absorbing material and the broad components that we have detected originate in the outer, and therefore slower, part of the BLR. In this case a trend should be visible where the largest FWHM are detected in the less X-ray absorbed and/or less reddened objects. As shown in Figure \ref{fig:FW_NH}, where we plot the FWHM as a function of $N_{\rm H}$ and as a function of the extinction $A_{\rm V}$, we do not find such trends in our AGN2 sample. 
The extinction $A_{\rm V}$ has been estimated in 8 AGN2 using either the BLR Paschen and Balmer line ratios (when available), assuming a Milky Way reddening law \citep{allen76} and R$_V$=3.1, or the values derived  by \citet{burtscher15} using a ``dust color'' method (see Table \ref{table1}).
The measured $A_{\rm V}$-$N_{\rm H}$ distribution of our AGN2 sample is typical of the AGN2 population \citep[see e.g.][Fig. 3]{burtscher16}.

\begin{figure}
	\centering
	\includegraphics[width=\columnwidth]{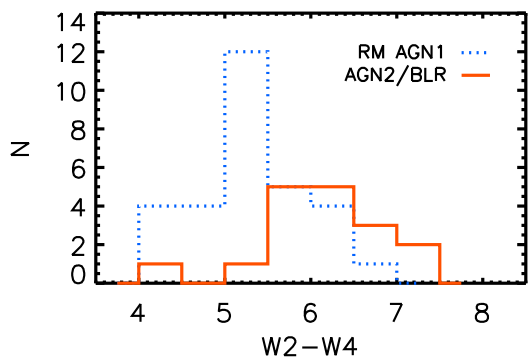}
	\caption{Distribution of the W2-W4 MIR color of our sample of  AGN2/BLR (red continuous line) and of the control sample of AGN1 whose \mbh\ have been measured via RM (blue dotted line).}
	\label{fig:HistW2W4}
\end{figure}

The presence of dust inside the BLR is unlikely as it would disagree with the results obtained through dust RM at K-band wavelength, that the time lag of the inner radius of the torus is three to four times longer than the H$\beta$ time lag \citep[][and references therein]{burtscher13, vazquez15}. The dust, instead, is consistent with being originated at the distance of  the predicted graphite sublimation radius  \citep{netzer93,netzer15}. This implies that the BLR is indeed bound by the dust distribution, as also observationally confirmed by \citet[][but see \citet{czerny11} for a different scenario in which the BLR is originated in regions where dust could co-exist]{landt14}.

We can therefore conclude that, once a BLR component is detected, the dusty region \citep[maybe clumpy, according  to recent studies;][and references therein]{marinucci16}  that surrounds the AGN should have been completely penetrated. 
Note that a possible consequence of dust absorption could be the reduction of the line intensity \citep[but not of the BLR FWHM; see also][]{kim15}. This could affect the estimation of the BH mass, if the line luminosity were to be used, as it is the case of some virial relations for type 1 AGN.
Therefore, in the case of obscured AGN, instead than the Optical/NIR line luminosities it is better to use the hard X-ray luminosities as a proxy of the BLR radius in the virial relations for the estimation of the BH mass. 

Another possible bias could be due to the fact that in the standard unification model obscured AGN are viewed, on average, at larger angles relative to the accretion disk axis than unobscured AGN. Therefore the estimation of the BLR FWHM could be affected by projection effects.
Indeed the dependence of BLR line widths in term of orientation is observationally well known  \citep{wills86}. Recently, \citet{bisogni17}, using the  EW of the [OIII] $\lambda$5007 \AA\ line as an inclination indicator, have found that the more inclined AGN1 have on average larger FWHM of the BLR H$\beta$  line. These results match with those of \citet{pancoast14}, who found that the virial {\it f} factor decreases with increasing inclination \citep[see also][]{risaliti11}.
A smaller {\it f} factor is needed to compensate the increasing  broadening of the FWHM with the
 inclination of the observed emission line width. These results have been interpreted as a hint at a possible disc-like shape for the BLR \citep{bisogni17}.

However, the  EW of the [OIII] line is not a good indicator of the orientation for AGN2, as the AGN continuum component is suppressed and overwhelmed by the host galaxy contribution. Therefore, following \citet{rose15}, we have used the W2-W4 (4.6-22 $\mu$m) mid-IR color, as measured by the {\it Wide-Field Infrared Survey Explorer} \citep[WISE,][]{wright10},
to roughly estimate the orientation of our samples of AGN. In Table \ref{table1} the W2-W4 colors of the AGN2/BLR sample are listed.
The more inclined AGN2 should show redder W2-W4 colors than AGN1 because the hottest dust emission of the inner regions of the torus should be less visible as the inclination increases \citep[][and references therein]{rose15}. 
Our AGN2/BLR show a red W2-W4 distribution, typical of the AGN2 population, having an average  $<$W2-W4$>$=6.2 with a (1$\sigma$) spread of 0.8, while the AGN1 control sample has  $<$W2-W4$>$=5.2 with a (1$\sigma$) spread of 0.7
\citep[see Fig. \ref{fig:HistW2W4} and, for comparison, Fig. 2 in][]{rose15}. If the common luminosity bin 42.5$<$logL$_{\rm 14-195}$/${\rm erg~ s^{-1}}$$<$44.5
is considered, AGN2/BLR (AGN1) have $<$W2-W4$>$=6.4 (5.4) with a (1$\sigma$) spread of 0.6 (0.7).

We can conclude that the smaller BLR FWHM and lower \mbh\ of our sample of AGN2 with respect to the AGN1 of the same  luminosity should not be ascribed to orientation effects. Indeed the larger inclination of the AGN2 should cause larger FWHM to be observed, and even smaller {\it f} factors to be used. Therefore, following \citet{pancoast14}, even smaller \mbh\ could be derived for the AGN2/BLR sample than those obtained in this work where a common $f=4.31$ factor, for both AGN1 and AGN2, is assumed.

\section{Conclusions}  

Determining the distribution of \mbh\ of AGN is of paramount importance in order to understand the AGN phenomenon.

Using  deep NIR spectroscopic observations we have detected faint BLR components in a sample of 17 AGN2  \citep{onori16a}
drawn from the 14-195 keV X-ray {\it Swift}/BAT 70-month catalogue, which is free of biases against absorbed sources up to log$N_{\rm H}\lesssim 24~ {\rm cm}^{-2}$ \citep{burlon11}.
No dependence 
was found of the BLR detection success rate, or FWHM of the lines, on either the X-ray or NIR fluxes, or on orientation angle of the host galaxy, on the hydrogen column density and on the  extinction.

In this work we have found that  the average AGN2 FWHM of the BLR is $\sim 0.25$ dex smaller than  measured in a control sample of AGN1 having the same average X-ray intrinsic luminosity.
Using new virial relations calibrated by \citet{Ricci16b}, which are based on the FWHM of the most relevant BLR NIR emission lines and the intrinsic hard X-ray  luminosity, we have measured the \mbh\ of the AGN2 in our sample.
The \mbh\ of the AGN2
are on average $\sim 0.8$ dex smaller than  measured in the control sample of AGN1. If AGN1 and AGN2 of the same luminosity log(L$_{\rm 14-195}$ / erg s$^{-1}$) $\sim$ 43.5  are compared, AGN2 have 0.5 dex smaller BH masses than the AGN1. 

Our findings are based on small samples and more observations are needed for more robust statistical grounds.
However, at face value our result disagrees with the standard AGN unification scenarios in which AGN1 and AGN2 are the same objects observed along different viewing angles with respect to  a toroidal absorbing material. 

Our findings could instead fit in an evolutionary scenario \citep[see e.g.][]{hopkins05} in which AGN2 represents the preceding stage of a type 1 AGN. In this picture,  AGN2 are dust enshrouded ``buried'', low mass (\mbh $\rm \lesssim 10^{7.5} M_{\odot}$) BH which accrete at high Eddington ratios. When feedback from the accretion drives away the obscuring material a window is created in which the AGN is seen as an optical type 1. In this evolutionary path highly accreting red quasars, as those observed by \citet{kim15b}, could belong to an intermediate population. Eventually, the activity ends when the accretion rate drops below that required to maintain the typical AGN luminosities.

\section*{Acknowledgments}
We thank the referee for  her/his very useful comments which improved the manuscript. We also thank Robert Antonucci, Ryan Hickox and Cesare Perola for useful discussions.
Based on observations made with ESO telescopes at the Paranal Observatory and the Large Binocular Telescope (LBT) at Mt. Graham, Arizona. 
We acknowledge funding from PRIN-MIUR 2010 and from PRIN-INAF 2011.
MB acknowledges support from the FP7 Career Integration Grant ``eEASy'' (CIG 321913). RM acknowledges support from the ERC Advanced Grant 695671
``QUENCH'' and from the Science and Technology Facilities Council (STFC).

\bibliographystyle{mnras} 
\bibliography{mybib} 

\label{lastpage}

\end{document}